\documentclass[authorversion,sigconf]{acmart}

\AtBeginDocument{%
  }

\copyrightyear{2025}
\acmYear{2025}
\setcopyright{cc}
\setcctype{by}
\acmConference[SC Workshops '25]{Workshops of the International Conference for High Performance Computing, Networking, Storage and Analysis}{November 16--21, 2025}{St Louis, MO, USA}
\acmBooktitle{Workshops of the International Conference for High Performance Computing, Networking, Storage and Analysis (SC Workshops '25), November 16--21, 2025, St Louis, MO, USA}
\acmDOI{10.1145/3731599.3767490}
\acmISBN{979-8-4007-1871-7/2025/11}

\usepackage{amsmath,amsthm}
\usepackage{listings}
\usepackage{sc25repro}

\usepackage[utf8]{inputenc}
\usepackage{physics}
\usepackage{amsbsy}
\usepackage{amsfonts}
\usepackage{bm}
\usepackage{mathtools}
\usepackage{float}
\usepackage{xfrac}

\usepackage{subcaption}

\hypersetup{colorlinks=true, citecolor=blue, linkcolor=blue, urlcolor=blue}
\usepackage{cleveref}

\theoremstyle{remark}
\newtheorem{remark}{Remark}

  \definecolor{ICES}{RGB}{94, 156, 174}
  \definecolor{ORANGE}{RGB}{191, 87, 0}
  \definecolor{RED}{RGB}{190, 30, 49}
  \definecolor{SUN}{RGB}{227, 81, 51}
  \definecolor{GREEN}{RGB}{0, 171, 86}
  \definecolor{BLUE}{RGB}{11, 78, 179}
  \definecolor{BROWN}{RGB}{122, 80, 40}
  \definecolor{GREY}{RGB}{50, 50, 50}
  \definecolor{TEAL}{RGB}{0, 160, 176}
\definecolor{OLIVE}{RGB}{93, 150, 72}

\newcommand{\be}{\begin{equation}}
\newcommand{\ee}{\end{equation}}

\newcommand{\bb}[1]{\mathbb{#1}}
\newcommand{\mc}[1]{\mathcal{#1}}

\newcommand{\p}{\partial}

\newcommand{\ptomat}{\mathbf{F}}

\newcommand{\invnoise}{\Gamma_{\hskip -1pt \text{noise}}^{-1}}

\newcommand{\bfinvnoise}{\mathbf{\Gamma}_{\hskip -1pt \text{noise}}^{-1}}
\newcommand{\bfinvpriorcov}{\mathbf{\Gamma}_{\hskip -1pt \text{prior}}^{-1}}

\newcommand{\noise}{\Gamma_{\hskip -1pt \text{noise}}}
\newcommand{\priorcov}{\Gamma_{\hskip -1pt \text{prior}}}
\newcommand{\postcov}{\Gamma_{\hskip -1pt \text{post}}}

\newcommand{\bfpostcov}{\mathbf{\Gamma}_{\hskip -1pt \text{post}}}

\newcommand{\pilike}{\pi_{\text{like}}}
\newcommand{\muprior}{\mu_{\text{prior}}}
\newcommand{\mupost}{\mu_{\text{post}}}
\newcommand{\mmap}{m_{\text{map}}}

\newcommand{\blocktoeptxt}{F}
\newcommand{\blocktoep}{\mathbf{\blocktoeptxt}}
\newcommand{\paramtxt}{m}
\newcommand{\datatxt}{d}
\newcommand{\paramvec}{\vb{\paramtxt}}
\newcommand{\datavec}{\vb{\datatxt}}
\newcommand{\bes}{\begin{equation*}}
\newcommand{\ees}{\end{equation*}}

\newcommand{\numtime}{N_t}
\newcommand{\numparam}{N_m}
\newcommand{\numdata}{N_d}

\crefname{subsection}{Section}{Sections}
\crefname{equation}{}{}

\begin{document}

\title[Mixed-Precision Performance Portability for FFTMatvec]{Mixed-Precision Performance Portability of\\FFT-Based GPU-Accelerated Algorithms for \\Block-Triangular Toeplitz Matrices}

\author{Sreeram Venkat}
\email{srvenkat@utexas.edu}
\orcid{0000-0001-8622-1063}
\affiliation{%
  \institution{Oden Institute, The University of Texas at Austin}
  \institution{Advanced Micro Devices, Inc.}
  \city{Austin}
  \state{Texas}
  \country{USA}
}

\author{Kasia {\'S}wirydowicz}
\email{kasia.swirydowicz@amd.com}
\affiliation{%
  \institution{Advanced Micro Devices, Inc.}
  \city{Austin}
  \state{Texas}
  \country{USA}
}
\author{Noah Wolfe}
\email{noh.wolfe@amd.com}
\affiliation{%
  \institution{Advanced Micro Devices, Inc.}
  \city{Austin}
  \state{Texas}
  \country{USA}
}
\author{Omar Ghattas}
\email{omar@oden.utexas.edu}
\affiliation{%
  \institution{Oden Institute, Walker Department of Mechanical Engineering}
  \institution{The University of Texas at Austin}
  \city{Austin}
  \state{Texas}
  \country{USA}
}

\begin{abstract}
The hardware diversity in leadership-class computing facilities, alongside the immense performance boosts from today's GPUs when computing in lower precision, incentivizes scientific HPC workflows to adopt mixed-precision algorithms and performance portability models. We present an on-the-fly framework using \texttt{hipify} for performance portability and apply it to FFTMatvec---an HPC application that computes matrix-vector products with block-triangular Toeplitz matrices. Our approach enables FFTMatvec, initially a CUDA-only application, to run seamlessly on AMD GPUs with excellent performance. Performance optimizations for AMD GPUs are integrated into the open-source rocBLAS library, keeping the application code unchanged. We then present a dynamic mixed-precision framework for FFTMatvec; a Pareto front analysis determines the optimal mixed-precision configuration for a desired error tolerance. 
Results are shown for AMD Instinct\texttrademark \ MI250X, MI300X, and the newly launched MI355X GPUs. 
The performance-portable, mixed-precision FFTMatvec is scaled to 4,096 GPUs on the OLCF \textit{Frontier} supercomputer.

\end{abstract}

\renewcommand{\shortauthors}{Venkat et al.}



\maketitle

\section{Introduction}\label{sec:introduction}

As the artificial intelligence (AI) market continues to drive GPU technology, hardware advancements are largely focused on accelerating lower precision computing. As a result, GPUs such as the AMD\footnote{AMD, the AMD Arrow logo, and combinations thereof are trademarks of 
Advanced Micro Devices, Inc.} Instinct\texttrademark \ MI355X and NVIDIA B200 have much higher peak throughputs for single (FP32) and half (FP16) precision workloads than for double (FP64) precision workloads. In addition, many consumer-grade GPUs have limited or no native FP64 support at all, and resort to emulation for double-precision calculations~\cite{goddeke2007performance}. It is important that traditional scientific high performance computing (HPC) workflows and algorithms are poised to leverage the advancements and trends in hardware. The prevailing methodology for this is to identify the computational portions of the scientific workflows that can be executed in lower precision while maintaining a satisfactory level of accuracy in the final result. Examples of this methodology include iterative refinement in solving linear systems~\cite{carson2018accelerating},  strategies to lower the precision of various components of linear solvers (i.e., preconditioner or matrix-vector product), see, for instance,~\cite{buttari2007mixed, guo2025adaptive}, and the Ozaki scheme for matrix multiplication~\cite{buttari2007mixed,ozaki2025ozaki}. Techniques for inserting mixed precision into scientific workloads have been in the spotlight for the last decade: see~\cite{abdelfattah2021survey,kashi2024mixed} and the references therein. By utilizing lower precision to compute intermediate results and switching to higher precision for final accumulations or calculation of residuals, these mixed-precision algorithms provide significant speedups over their traditional, double-precision counterparts. Iterative methods maintain the desired level of accuracy in the final result by taking more iterations, though the cost of each (lower-precision) iteration is reduced. Similarly, the Ozaki scheme incurs increased memory utilization to store decomposed, lower-precision matrices. However, by doing so, the Ozaki scheme can leverage matrix/tensor cores on GPUs that have much higher throughputs for lower-precision workloads. 

Another important consideration in the field of HPC application development is that of performance portability. Leadership-class computing facilities remain diversified across various vendors---with the Oak Ridge Leadership Computing Facility's \textit{Frontier} and Lawrence Livermore National Laboratory's \textit{El Capitan} systems using AMD hardware, the National Energy Research Scientific Computing Center's \textit{Perlmutter} and newly announced \textit{Doudna} systems using NVIDIA hardware, and the Argonne Leadership Computing Facility's \textit{Aurora} system using Intel hardware. In addition, many new specialized AI chips from companies such as Cerebras are also being considered for scientific HPC workflows~\cite{ltaief2023scaling}. The Exascale Computing Project from the DOE led to the development of many performance portability frameworks, including Kokkos and Raja~\cite{trott2021kokkos,beckingsale2019raja}. The SYCL C++ programming model from the Khronos group, the OCCA portability framework developed at Rice University and supported by Argonne National Laboratory, and the Legion programming system from Stanford~\cite{reyes2016sycl,medina2014occa,bauer2012legion} are further alternatives for performance portability. These portability frameworks provide abstraction layers that allow users to implement parallel algorithms in a vendor-agnostic manner. They can be particularly useful when developing new applications or software libraries~\cite{paunovic2024utilizing,deakin2020evaluating}. However, it is often the case that developers have an existing application written in a vendor-specific language---usually CUDA---but would like to run on hardware from a different vendor. For these cases, the aforementioned performance portability frameworks are less directly applicable. Integrating these performance portability frameworks into an application usually requires significant refactoring of the codebase. Additionally these frameworks often involve many advanced C++ structures (lambdas, templating, etc.) that make it more difficult to read and understand. Moreover, these frameworks generally follow a programming paradigm that is syntactically very different from CUDA. These factors pose a substantial obstacle to the many application developers who would much prefer to maintain a single CUDA source code but still be able to run their software on hardware from different vendors. Many developers were introduced to GPU programming through CUDA, and many universities still teach CUDA as the only GPU programming model in computer science curricula~\cite{Morgan2021CUDAMoat}.

For these situations, the HIP programming model aims to provide a solution.\footnote{\url{https://rocm.docs.amd.com/projects/HIP/en/latest/}} By mirroring the CUDA programming paradigm, it enables a relatively simple mapping of source code. For the same reason, it is also a much more readable and understandable model for developers used to CUDA. While HIP currently only natively supports AMD and NVIDIA backends, there are ongoing third-party efforts to support HIP on Intel hardware as well.\footnote{See, for example: \url{https://github.com/CHIP-SPV/chipStar}} As a result, HIP provides cross-platform capability that spans the majority of today's HPC hardware providers. The AI hardware companies, on the other hand, are a wholly different case; they come with specialized software stacks and are not generally supported by the other performance portability models (Kokkos, Legion, etc.) either. There are projects such as SCALE\footnote{\url{https://docs.scale-lang.com/nonfree-unstable/}} which aim to compile CUDA source code directly into AMD binaries; however, these are still under development and not yet ready for testing on production-level applications.

While HIP's cross-platform compatibility and similarity to CUDA are attractive to HPC application developers~\cite{alexander2021survey}, the issue of code translation remains: what is the best way to deal with an existing CUDA source code? Some developers have taken the route of maintaining custom header files that use preprocesser definitions to ``toggle'' between CUDA and HIP at compile time. This method has the advantage of being lightweight but has the disadvantage that every time new functionality is added to the CUDA code, the header file has to be updated manually. The \texttt{hipify}\footnote{\url{https://rocm.docs.amd.com/projects/HIPIFY/en/latest/index.html}} tool from AMD provides an answer to the translation by converting CUDA source code to HIP. When combined with a build system such as CMake, \texttt{hipify} can be configured to run ``on-the-fly.'' Thus, the only maintained source code is in pure CUDA; this source is \textit{hipified} at compile time and then compiled into an executable that can run on AMD GPUs. Compiling for NVIDIA GPUs remains the same as always---no \textit{hipification} needed. This process
provides a viable and sustainable answer to the problem of code translation. 

A question may arise on how to handle the cases where a certain functionality is supported by a CUDA library but is not present in the corresponding ROCm\texttrademark\ or HIP library. This problem is not unique to HIP; any of the previously discussed performance portability frameworks would also encounter this issue. For these cases, the \texttt{hipify} tool can be directed via preprocessor directives to use any custom implementation of the same functionality or else throw a "Not Supported" error. Furthermore, as the ROCm and HIP libraries are open source,\footnote{\url{https://github.com/ROCm/rocm-libraries}} it is possible to integrate a custom implementation of a functionality into a local build of the library and then point \texttt{hipify} and the HIP compiler to use that via the build system. This allows for a flexible environment where custom implementations that improve performance and portability can be seamlessly merged into the workflow. 

In this paper, we will present performance portability via \texttt{hipify} on-the-fly
and mixed-precision compute capability for an application that computes FFT-based matrix-vector products (``matvecs'') for block-triangular Toeplitz matrices~\cite{venkat2024fft}. These matrices are relevant in the context of Bayesian inverse problems for linear autonomous dynamical systems, where they enable fast Hessian actions. The algorithm for block-triangular Toeplitz matvecs described in that paper can be used to provide many orders of magnitude speedup over traditional methods of computing the Hessian action for these problems~\cite{henneking2025goal,henneking2025bell}. 

The paper is organized as follows. In~\Cref{sec:background}, we present an overview of the matvec algorithm and computational components involved. \Cref{sec:methods} discusses the performance portability of the algorithm, highlighting a special instance where custom functionality is integrated to produce significant performance improvements. \Cref{sec:methods} also discusses the implementation of mixed precision into the matvec algorithm. \Cref{sec:results} presents numerical results of the performance-portable, mixed-precision implementation. Finally,~\Cref{sect:conclusion} concludes the work.

\subsection{Related Work}
The \texttt{hipify} tool has been used in several large-scale HPC applications, including HACC, GROMACS, and LAMMPS~\cite{habib2013hacc,kondratyuk2021gpu,hagerty2023studying} to achieve cross-platform performance. Similarly, for AI applications, \texttt{hipify} has been used to convert existing CUDA backends for deep learning into HIP backends.\footnote{\url{https://github.com/ROCm/hipify_torch}} The exact integration scheme for \texttt{hipify} varies for each application; some convert once to a pure HIP source and then maintain that, while others use an approach similar to that of \texttt{hipify} on-the-fly
discussed previously. In all these cases, the vast majority of existing CUDA source code is automatically converted to HIP; developers may have to manually add support for any CUDA libraries or functionality lacking a HIP counterpart. In some cases, after \textit{hipification}, HIP kernels are tuned to achieve optimal performance on AMD GPUs~\cite{pall2014tackling,hagerty2023studying}. In this work, after presenting our dynamic \textit{hipification} framework, we will exemplify how this kernel tuning process can be integrated into the open-source ROCm or HIP libraries while leaving application source code unchanged. This approach contrasts with the one taken by many applications, which involves maintaining two sets of backend source files~\cite{pall2014tackling,kolev2020support}. 

In addition to performance portability, the use of mixed precision algorithms for general linear algebra (BLAS) routines and fast Fourier transforms (FFTs) has been widely studied~\cite{higham2022mixed,sorna2018optimizing}. There has also been work on mixed-precision algorithms for Toeplitz matrices; however, to our knowledge, the case for block-triangular Toeplitz matrices without a recursive Toeplitz structure has not been studied. This paper also analyzes the mixed-precision matvec algorithm for block-triangular Toeplitz matrices in the context of their application to Bayesian inverse problems. Using application-specific knowledge, such as the noise level present in the data and the numerical stability of subsequent computations in the application's workflow, a threshold for the acceptable error level in the mixed-precision algorithm can be determined. This enables a dynamic method for selecting which phases of the algorithm are computed in lower precision, thereby maximizing computational speedup while keeping the overall error level below the acceptable threshold.

\section{Background}\label{sec:background}

The matvec algorithm developed in~\cite{venkat2024fft} is applicable to general block-triangular Toeplitz matrices. These matrices can arise in several application contexts, including multi-channel signal processing, vector-autoregressive-moving-average models in econometrics, and inverse problems governed by linear autonomous dynamical systems~\cite{kailath1980linear,simpkins2012system,lutkepohl2013introduction,saad2003iterative}. In this section, we focus on the last of these applications and begin with an overview of the linear autonomous dynamical systems and Bayesian inverse problems. We then introduce relevant notational conventions that will be used in the remainder of the paper. Finally, we close with an outline of the FFT-based, GPU-accelerated matvec algorithm for block-triangular Toeplitz matrices. More details on the algorithm and its applications to inverse problems can be found in~\cite{venkat2024fft,henneking2025goal,henneking2025bell}.

\subsection{Linear Autonomous Dynamical Systems}
Dynamical systems refer to the broad class of systems whose evolution can be described by a function or rule. These systems can be used to describe everything from planetary motion to the spread of disease in a population. Many of the dynamical systems describing physical phenomena of interest are formulated as mathematical models expressed through Partial Differential Equations (PDEs). 

An \textit{autonomous} dynamical system is one whose evolution does not explicitly depend on the independent variable of the system. Most often, this independent variable is time; such systems are also called \textit{time-invariant} dynamical systems. A \textit{linear} autonomous dynamical system has the additional property that the mapping from input to output is a linear map. We consider linear time-invariant (LTI) systems of the form
\begin{equation}
\left\{
\begin{aligned}
	\frac{\partial u}{\partial t} &= \mc A u + \mc C m && \text{in } \Omega \times (0,T),  \\
	u &= u_0 && \text{in } \Omega \times \{ 0 \},  \\
	d &= \mc B u && \text{in } \Omega \times (0,T), \label{eq:LTI}
\end{aligned}
\right.
\end{equation}

with appropriate boundary conditions on the spatiotemporal domain $\p \Omega \times (0,T)$. In this formulation, $u(x,t)$ is the state variable with initial value $u_0(x)$; $m(x,t)$ is the parameter representing the source or forcing of the system and is independent of the state, and both $\mc A$ and $\mc C$ are time-invariant differential operators; $d(x,t)$ is the observable of the system, extracted from the state $u$ via a time-invariant observation operator $\mc B$. LTI systems of this form can be used to model heat transfer, diffusion, porous media flow, and wave propagation, where $m$ represents a source term.

The \textit{parameter-to-observable} (p2o) map $\mc F$ is defined by 
\begin{align}
    \mc F : m(x,t) \mapsto d(x,t),
\end{align}
via solution of the PDE~\cref{eq:LTI} using $m(x,t)$ as input and extraction of observations $d(x,t)$ from the state $u(x,t)$ as output. The map $\mc F$ is time invariant: $m(x,t+\tau) \mapsto d(x,t+\tau)$ is the same as the map $m(x,t) \mapsto d(x,t)$. The \textit{adjoint} p2o map $\mc F^*$ maps data $d(x,t)$ to parameters $m(x,t)$ by solving the adjoint system of PDEs corresponding to~\cref{eq:LTI}; it is also time invariant.

\subsection{Bayesian Inverse Problem}\label{sec:bip}
Given observations $d^{\text{obs}}(x,t)$ of the dynamical system~\cref{eq:LTI}, we want to infer the corresponding parameters $m(x,t)$ and quantify the uncertainty associated with this inference. This process is formalized as a \textit{Bayesian} inverse problem where the goal is to determine the \textit{posterior} measure $\mupost$ of the parameters given data. Bayes' theorem gives that 
\begin{align}
    \frac{\dd\mupost}{\dd \muprior} = \pilike\qty(d|m)\label{eq:bayes},
\end{align}
\sloppy where $\pilike(d|m)$ is the \textit{likelihood} distribution of the data given parameters, and $\muprior$ is the \textit{prior} distribution representing prior knowledge about the parameters. Assuming a Gaussian prior $m~\sim~\mathcal{N}(m_{\text{prior}}, \priorcov)$, Gaussian likelihood $\pilike\qty(m|d)~=~\exp\qty(-\tfrac{1}{2}\|\mathcal{F}m - d\|_{\invnoise}^2)$, linear p2o map $\mc F$, observations $d^{\text{obs}} = \mathcal{F}m + \nu$, and noise $\nu \sim\mathcal{N}(0,\noise)$, the posterior can be analytically expressed as $\mu_{\text{post}} = \mathcal{N}(m_{\text{map}}, \postcov)$. Thus, determining the posterior measure amounts to calculating $\mmap$ and $\postcov$~\cite{ghattas2021learning}. When formulated in the infinite-dimensional setting, the derivative in~\cref{eq:bayes} is the Radon-Nikodym derivative~\cite{stuart2010inverse}. However, for the remainder of the paper, we will deal only with the discretized versions of these problems.

\subsection{Notation and Problem Description}
Notationally, boldface will be used to denote the discrete versions of objects, while script will be used to denote the continuous (infinite-dimensional) versions of those objects. In the discrete notation,
\begin{align}
    \bfpostcov = \left(\mathbf{F}^* \bfinvnoise \mathbf{F} + \bfinvpriorcov \right)^{-1}\nonumber\\
    \vb{m}_{\text{map}}=\bfpostcov \left(\mathbf{F}^* \bfinvnoise \vb{d} + \bfinvpriorcov \vb{m}_{\text{prior}}\right).\label{eq:map-point}
\end{align}
Thus, the goal of the Bayesian Inverse Problem in this case is to solve for the \textit{maximum a posteriori} (MAP) point via the linear system in~\cref{eq:map-point}. Uncertainty can be quantified through the posterior covariance $\bfpostcov$. Traditional methods for solving this inverse problem using iterative solvers (e.g., conjugate gradient) and matrix-free actions of the \textit{Hessian} $\vb{H} \coloneqq \bfpostcov^{-1}$ are detailed in~\cite{ghattas2021learning}. For cases where the Hessian has a high effective rank, rendering an iterative solution of the linear system computationally intractable, novel algorithms have been developed in~\cite{henneking2025goal} and demonstrated on a large-scale example in~\cite{henneking2025bell}. Both methodologies for solving the inverse problem rely on actions of the p2o map $\blocktoep$ and its adjoint $\blocktoep^*$. The time invariant nature of $\mc F$ manifests in the discrete $\blocktoep$ being a \textit{block lower-triangular Toeplitz} matrix. Here, $\numparam$ is the number of spatial parameter points, $\numdata$ is the number of sensors ($\numdata \ll \numparam$), and $\numtime$ is the temporal dimension of parameters and observations ($\numtime \gg 1$). Then, the discrete parameters and observations are:
\begin{itemize}
	\item $\paramvec \in \bb R^{\numparam \numtime}$ with blocks $\paramvec_j \in \bb R^{\numparam}$, $j = 1,2, \ldots, \numtime$;
	\item $\datavec \in \bb R^{\numdata \numtime}$ with blocks $\datavec_i \in \bb R^{\numdata}$, $i = 1,2, \ldots, \numtime$.
\end{itemize}
The discrete p2o map is:
\bes
	\left[ \begin{array}{@{}c@{}}
	\datavec_1 \\[2pt]
	\datavec_2 \\[6pt]
	\datavec_3 \\[1pt]
	\vdots \\[3pt]
	\datavec_{\numtime}
	\end{array} \right]
	=
	\left[ \begin{array}{@{}ccccc@{}}
	\ptomat_{11} & \vb 0 & \vb 0 & \cdots & \vb 0 \\[2pt]
	\ptomat_{21} & \ptomat_{11} & \vb 0 & \cdots & \vb 0 \\
	\ptomat_{31} & \ptomat_{21} & \ptomat_{11} & \ddots & \vdots \\
	\vdots & \vdots & \ddots & \ddots & \vb 0 \\[2pt]
	\ptomat_{\numtime,1} & \ptomat_{\numtime-1,1} & \cdots & \ptomat_{21} & \ptomat_{11}
	\end{array} \right]
	\left[ \begin{array}{@{}c@{}}
	\paramvec_1 \\[2pt]
	\paramvec_2 \\[6pt]
	\paramvec_3 \\[1pt]
	\vdots \\[3pt]
	\paramvec_{\numtime}
	\end{array} \right] ,
	\label{eq:ShiftInvariance}
\ees
where $\blocktoep$ has block dimension $\numtime \times \numtime$, and $\blocktoep_{ij}~\in~\mathbb{R}^{\numdata \times \numparam}$.

\subsection{Matvec Algorithm}\label{sec:algorithm}
The block-triangular Toeplitz structure implies that only the first block column of $\blocktoep$ needs to be stored. Moreover, it can be computed via only $\numdata$ (number of sensors) adjoint PDE solutions. Furthermore, $\blocktoep$ can be embedded in a block circulant matrix---corresponding to a zero padding of the first block column. This block circulant matrix is block-diagonalized by the discrete Fourier transform \cite{gray2006toeplitz}; in Fourier space, the p2o matvec is simply a block-diagonal matvec. 

This structure of $\blocktoep$ motivates an FFT-based p2o matvec algorithm that can be implemented efficiently on multi-GPU clusters. The full algorithm is detailed in~\cite{venkat2024fft}; the main computational phases involved are\footnote{There are additional intermediate phases involving reordering of the vectors. These are purely memory operations, and we always compute them in the lowest possible precision given the compute precisions of the major phases adjacent to them.}
\begin{enumerate}
    \item Broadcast and add zero-padding to the input vector $\vb{m}$
    \item Compute FFT of the input vector $\vb{m} \mapsto \hat{\vb{m}}$
    \item Compute block-diagonal matvec in Fourier space (computed as a batched matrix-vector product) $\hat{\vb{m}}\mapsto\hat{\vb{d}}$
    \item Compute IFFT of the output vector $\hat{\vb{d}} \mapsto \vb{d}$
    \item Remove padding and compute reduction of the output vector $\vb{d}$
\end{enumerate}
It is important to note that all the phases operate on block vectors and matrices; all of the operations are batched (e.g., zero-pad each vector block, compute batched FFT, etc.). The algorithm for matvecs with $\vb{F}^*$ is identical except that in Phase 3, a conjugate transpose batched matrix-vector product is used (and the input/output vectors are switched).

In general, the FFTMatvec algorithm is designed to run on a 2D processor grid of shape $p_r\times p_c$, where $p\coloneqq p_rp_c$ is the total number of processors (GPUs). In Section 3.7 of~\cite{venkat2024fft}, an algorithm for communication-aware partitioning is described. This algorithm uses the problem size, number of available processors, and other system parameters to set the 2D process grid dimensions for FFTMatvec. For many applications running on a small to moderate number of GPUs ($\lesssim 512$), $p_r = 1$ and $p_c = p$ will be the optimal processor grid shape. In that case, the only nontrivial communication in the $\blocktoep$ matvec is the reduction in Phase 5, and the only nontrivial communication in the $\blocktoep^*$ matvec is the broadcast in Phase 1. When scaling to larger numbers of GPUs, especially across multiple racks of a machine, more than one row can be used in the processor grid to minimize the communication costs. 

An animated depiction of the full FFTMatvec algorithm is available on \href{https://youtu.be/hc81_WzGF_Q}{\color{red}{YouTube}}. In the subsequent sections, we will discuss the performance portability of this matvec algorithm and present a dynamic framework for using mixed precision in the matvec computation.

\section{Methods}\label{sec:methods}

In this section, we first discuss the performance portability of the matvec implementation using the \texttt{hipify} tool on-the-fly. As part of this discussion, we present an example of how custom performance optimizations can be integrated into the application. Afterwards, we describe the dynamic framework for mixed-precision computation of the matvecs. 

\subsection{Performance Portability via Hipify On-The-Fly}
The original, pure CUDA source code for the matvec implementation is available open-source at \url{https://github.com/s769/FFTMatvec} (FFTMatvec). Much of the implementation uses CUDA libraries---cuBLAS, cuFFT, cuTENSOR, and NCCL---and employs custom GPU kernels for operations such as zero-padding and unpadding. There are two versions of the \textit{hipify} tool provided as part of the ROCm software suite: \texttt{hipify-clang} and \texttt{hipify-perl}. The \texttt{hipify-clang} tool is a full-fledged translator that converts CUDA source code into an abstract syntax tree, traverses the tree using transformation matchers, and produces HIP source code. It is a well-supported compiler extension that also checks for correctness of the original CUDA code. On the other hand, \texttt{hipify-perl} is a more lightweight tool that uses regular expressions to translate CUDA source code directly into HIP; it is essentially an advanced find-and-replace tool. Both of these tools can be integrated into the build system to perform the conversion at compile time. For the FFTMatvec code, we used CMake functionality to create a build framework that, at compile time, calls the \textit{hipify} tool to convert the CUDA source into HIP source files that are placed in the build directory. Then, the HIP compiler compiles these \textit{hipified} source files into executables that can run on AMD GPUs. The \textit{hipification} can be toggled by setting a CMake variable; when it is off, the NVIDIA binaries are built from the CUDA source code as usual. Thus, the only source code that needs to be maintained is the original CUDA code; whenever a change is made in the CUDA source, recompilation automatically triggers re-\textit{hipification} of the modified source files. 

For \textit{hipification}, both \texttt{hipify-clang} and \texttt{hipify-perl} were tested. For the FFTMatvec application, it is enough to use the lightweight \texttt{hipify-perl}; this avoids the need for an otherwise useless CUDA installation on an AMD machine. 

The automated \textit{hipification} successfully converted almost all of the CUDA source. However, there were some functionalities from cuTENSOR that were not yet supported in hipTensor. Namely, these were the permutation functionalities for complex double-precision datatypes present in cuTENSOR (v2). The corresponding functionalities for hipTensor are planned in an upcoming release; for the time being, the same functionality can be implemented via a custom GPU kernel. The algorithm used for this kernel is a modification of the one developed in~\cite{jodra2015efficient} to avoid overflowing the maximum number of grid blocks that can be launched in the $y$ and $z$ dimensions. This kernel is comparable in performance to the original cuTENSOR permutation and is only used in the setup phase of the computation; it is not a part of the performance critical portion. Moreover, by removing the cuTENSOR (v2) dependency from the original CUDA code itself, this approach further increases portability. Many NVIDIA systems are equipped with the cuTENSOR (v1) library, and originally, one had to manually install the v2 library before building the FFTMatvec application. 

\subsubsection{Performance Optimization for AMD GPUs}\label{sec:perf-opt}
After removing the cuTENSOR dependency, the \textit{hipification} process enables the CUDA application to seamlessly run on AMD GPUs. However, when running performance tests, we observed a performance reduction in matvecs involving the adjoint $\blocktoep^*$ matrix when compared to matvecs involving the $\blocktoep$ matrix. After running a timing analysis and profiling the code using \texttt{rocprofv3}, we found that the strided batched GEMV kernel in rocBLAS (SBGEMV)---see~\Cref{sec:algorithm}---was attaining a much lower memory bandwidth when running in conjugate transpose mode as compared to the non-transpose mode. Since the rocBLAS library is open-source (available at \url{https://github.com/ROCm/rocm-libraries/tree/develop/projects/rocblas}), we were able to easily diagnose the issue. 

The crux of the problem is that for applications in inverse problems, the number of sensors $N_d$ is generally much less than the number of spatial parameters $N_m$. This is because each sensor installation usually involves some sort of cost, while the spatial parameter dimension can be arbitrarily large for high-order PDE discretizations over large spatial domains. The SBGEMV in Phase 3 of the matvec algorithm in~\Cref{sec:algorithm} operates on a batch of $N_t+1$ matrices of size $N_d \times N_m$. The matrix elements are complex numbers since this operation is in Fourier space. When $N_d \ll N_m$, these matrices are short and wide; the rocBLAS kernel that is selected for the non-transpose matvec is launched with grid dimensions of $\texttt{ceil}(N_d/64)~\times~1~\times~(N_t+1)$. On the other hand, the conjugate transpose matvec kernel is launched with grid dimensions of $N_m~\times~1~\times~(N_t+1)$. The batching over $N_t +1$ is handled by the third grid dimension, so it is enough to consider the first grid dimension to analyze this problem. In the non-transpose case, each gridblock computes several dot products of size $N_m$; in the conjugate transpose case, each gridblock computes a single dot product of size $N_d$. Thus, when $N_d << N_m$, the conjugate transpose kernel launches many gridblocks that each has very little work. This results in increased launch overheads and decreased memory bandwidth.

To address this issue, we developed an alternative kernel to handle SBGEMVs for batches of $m\times n$ matrices where $m < n$. Though our application only utilizes complex dataypes, we developed kernels that handle both real and complex matvecs in single and double precision. Additionally, we handled both the regular transpose and conjugate transpose cases. The algorithm employs a tiling approach where the kernel gridblocks tile the columns of each matrix in the batch. Each gridblock itself comprises a 2D set of threads; thus, each gridblock is responsible for computing a \textit{chunk} of elements of the output vector. This is in contrast to the original rocBLAS kernel, where each threadblock computes a single element of the output vector. In addition, vectorized data loads and pipelining are used to achieve higher memory bandwidth in the kernel. In a single instruction, a maximum of 16 bytes can be read or written by a thread; using vectorized datatypes such as \texttt{float4} and \texttt{double2} allow multiple elements to be fetched by each thread in a single instruction while maintaining coalesced memory access. Additionally, pipelining the read, compute, and write instructions overlaps the memory and compute instructions and hides the read/write latencies behind computations. Finally, warp---or wave---shuffles are used to do the reductions for computing the dot products in the SBGEMV. 

The tiling and gridblock size parameters, as well as the amount of data vectorization and pipelining, vary for each datatype (\texttt{float}, \texttt{double}, \texttt{complex\_float}, and \texttt{complex\_double}). As a result, a separate kernel is used for each datatype; this is hidden from the user by a templated host-side dispatch function. These algorithmic modifications allow the kernel to attain much higher memory bandwidths (see~\Cref{sec:results}) and thus solve the performance issue for the $\blocktoep^*$ matvecs.

When implementing tuned kernels for specific architectures, a potential implementation strategy could be to use preprocessor directives that guide the compiler to select a specific kernel based on the device hardware type (AMD or NVIDIA). However, in this case, a much simpler solution is to clone the open-source rocBLAS library, insert this custom kernel into the rocBLAS host dispatcher, build the library, and link the application against it. With this method, the application code is \textbf{completely unchanged} and the performance improvements are automatically added to the application. As the ROCm\texttrademark\ software suite continues to evolve, this approach will work for an increasing number of applications, enabling robust performance portability optimization without increasing code complexity. In addition, user performance optimizations can eventually be merged into the core rocBLAS library via pull request---we have merged this short and wide (conjugate) transpose SBGEMV kernel into the rocBLAS development branch. This process prevents other application developers from having to ``reinvent the wheel'' and creates a collaborative software ecosystem where algorithmic innovations can thrive.

\subsection{Dynamic Mixed-Precision Framework}\label{sec:mp}
In addition to performance portability of the FFTMatvec application, we also developed a framework for dynamically applying mixed-precision computation in the matvec. We note that the input and output of the matvecs will be assumed to be in double precision. This is primarily due to the fact that when applying these accelerated p2o matvec algorithms to inverse problems, a dense, data-space Hessian matrix is formed by taking actions of $\blocktoep$ and $\blocktoep^*$~\cite{henneking2025goal}. For discretizations of ill-posed inverse problems where sparse, noisy data have to inform many parameters, the conditioning of this dense matrix is often poor. Thus, further computations are carried out in double precision to avoid accuracy issues due to roundoff error. The lowest precision that is used in the computation will be single (FP32); while half-precision performance can be extremely high on the latest GPU architectures, software support for half-precision linear algebra and FFT routines---especially those involving complex numbers---is sparse.

The general framework is based on the decomposition of the matvec algorithm into the five phases described in~\Cref{sec:algorithm}. For each of these phases, the computation can be performed in either single (FP32) or double (FP64) precision. The compute precisions of each phase of the algorithm can be set through a precision configuration \texttt{struct} that is passed as an argument when creating the matrix. The compute precision of each phase also determines the precisions of the matrix and input/output vectors at that phase. This precision configuration can be set at runtime, allowing for dynamic testing of various mixed-precision configurations. 

The current working precision is tracked throughout the matvec computation, and it always begins and ends in double precision in accordance with the earlier discussion. If a given phase of the matvec algorithm needs to be performed in a precision different from the current working precision, a cast is performed. At all possible points, the casting kernels are fused with any nearby memory operations (zero-padding, unpadding, etc.) to reduce kernel launch latencies associated with launching multiple small kernels. In addition, all memory operations---zero-padding, unpadding, vector reorderings---are performed in the lowest possible precision among the compute precisions of adjacent phases. The matrix setup phase is performed in double precision as it is a one-time operation that is not performance critical.

This dynamic mixed-precision framework for the matvec algorithm allows us to determine the ideal precision configuration for a given application. The Pareto front~\cite{deb2001multi} can be used to quantify this idea: for a set error tolerance, choose the precision configuration that gives the greatest performance improvement while keeping the error below that tolerance. The error tolerance can be determined based on the application; the data vector $\vb{d}$ that contains observations from the sensors will have some associated measurement precision or tolerance. In addition, there will be some amount of noise in the data. Thus, the sensor tolerance and assumed noise level can be used to set the error tolerance for the mixed-precision matvec. 

\begin{remark}\label{rem:why} Considering the matvec computation times for very large-scale inverse problem applications is $\order{10}$ms~\cite{henneking2025bell}, a question may arise as to why we would want to further speed up the matvec through mixed-precision computations. In short, the answer lies in the fact that while solving a single inverse problem real-time only requires a handful of block-triangular Toeplitz matvecs~\cite{henneking2025goal}, we are also interested in tackling additional ``outer-loop'' problems. One very important outer-loop problem is that of optimal sensor placement. In the literature, this is often done by choosing a sensor placement that maximizes the expected information gain, measured by the Kullback-Leibler divergence between the prior and posterior~\cite{AlexanderianPetraStadlerEtAl14}. Since we have a linear inverse problem with a Gaussian prior and posterior, the KL-divergence has a closed form that depends on the sensor locations~\cite{AlexanderianPetraStadlerEtAl14,AlexanderianGloorGhattas16}. This can then be used to solve for optimal sensor locations. One option is to select from a subset of viable locations and use sparsity-promoting regularization~\cite{AlexanderianPetraStadlerEtAl16} or use a greedy algorithm~\cite{wu2023fast}. All of these methods will require re-assembling the dense, data-space Hessian matrix that requires $N_dN_t$ actions of $\blocktoep$ and $\blocktoep^*$ ($\mathcal{O}(10^5)$ for large-scale problems~\cite{henneking2025bell}). So, when testing many sensor configurations, any performance improvements in the matvec algorithm will be made much more relevant in these computations.
\end{remark}

\subsubsection{Numerical Error Analysis}\label{sec:error}
In this section, we present a theoretical analysis of the numerical error in the mixed-precision FFTMatvec algorithm. We compute the errors in each phase of the algorithm and propagate them to the subsequent phases. The error analysis is computed to first order, following standard methodologies~\cite{higham2002accuracy}.

Notationally, begin with an initial vector $\vb{v}_0$, which is an 
\textit{exact double-precision floating-point vector} representing the starting data.

From this, define two sequences:
\begin{enumerate}
    \item The \textbf{true vector} at subsequent stages, $\vb{v}_i$, is the ideal, 
    infinite-precision result of applying the mathematical operation in Phase $i$ 
    to the \emph{previous true vector} $\vb{v}_{i-1}$. This sequence, 
    $\{\vb{v}_1, \vb{v}_2, \dots\}$, represents the perfect mathematical path the data would follow if no rounding errors ever occurred.
    \item The \textbf{computed vector}, denoted by $\vb{v}'_i$, is the actual 
    floating-point result stored in the machine after Phase $i$. It is obtained 
    by applying the finite-precision operation in Phase $i$ to the \emph{previously computed vector} 
    $\vb{v}'_{i-1}$. This sequence, $\{\vb{v}'_1,\vb{v}'_2, \dots\}$, 
    represents the path the data actually takes.
\end{enumerate}

The error at Phase $i$ measures the total deviation of the actual 
computational path from the ideal mathematical path, and is defined as:
\begin{equation}
    \delta\vb{v}_i = \vb{v}'_i - \vb{v}_i.
\end{equation}

Additionally, $\epsilon_i$ denotes the machine epsilon corresponding to the precision that is used to compute Phase $i$. That is, $\epsilon_i = \epsilon_s\approx 10^{-7}$ for single-precision phases and $\epsilon_i = \epsilon_d \approx10^{-16}$ for double-precision phases. Finally, $c_i$ denotes any $\order{1}$ algorithm-dependent constants for Phase $i$.

Recall that FFTMatvec uses a 2D processor grid of shape $p_r\times p_c$, where $p\coloneqq p_rp_c$ is the total number of processors (GPUs) in general. The processor grid dimensions affect the error analysis; the analysis below reflects this.

\paragraph{Initial state (Phase 0)} The input vector $\vb{v}_0$ is assumed to have an exact floating-point representation. A batched FFT of $\vb{F}$ is precomputed during the initialization of FFTMatvec to transform the block-Toeplitz structure into a block-diagonal structure. This computation is always done in double precision; the error in the computed $\hat{\vb{F}}'$ is $\delta\hat{\vb{F}}$. Moreover, the relative error $\|\delta\hat{\vb{F}}\|/\|\hat{\vb{F}}\| \leq c_F\epsilon_d\log_2(2N_t)$~\cite{vanloan1992computational}.

\paragraph{Broadcast and Zero-Padding} These are purely memory operations. As a result, the error will be zero if computed in double precision. If computed in single precision, the error will be bounded as $\|\delta\vb{v}_1\|\leq \epsilon_s\|\vb{v}_0\|$. These cases are combined by defining $c_1 \coloneqq 0$ if Phase 1 is computed in double precision and $c_1 \coloneqq 1$ if Phase 1 is computed in single precision. Then, 
$$\|\delta\vb{v}_1\|\leq c_1\epsilon_1\|\vb{v}_0\|$$

\paragraph{Batched FFT} Each block in the vector has size $2N_t$ (after padding). The error after this phase is $\delta \vb{v}_2 = \text{FFT}(\delta\vb{v}_1) + (\text{error in FFT}(\vb{v}_1))$. Now, the norm of the FFT operator is $\sqrt{2N_t}$. Using the standard FFT error result from~\cite{vanloan1992computational} gives
$$\|\delta\vb{v}_2\|\leq \sqrt{2N_t}\|\delta\vb{v}_1\| + c_2\epsilon_2\sqrt{2N_t}\log_2(2N_t)\|\vb{v}_1\|.$$

\paragraph{SBGEMV} This phase is formally a block-diagonal matvec with $\hat{\vb{F}}$ or $\hat{\vb{F}}^*$.\footnote{When computing on multiple processes, the matvec is with the \textit{local portion} of $\hat{\vb{F}}$ or $\hat{\vb{F}}^*$. Since this is also a block diagonal matrix, the analysis remains the same.} Without loss of generality, we show the analysis for the $\hat{\vb{F}}$ case. To first order, $\delta \vb{v}_3 = \hat{\vb{F}}\delta\vb{v}_2 + \delta\hat{\vb{F}}\vb{v}_2 + (\text{error in } \hat{\vb{F}}\vb{v}_2)$. Each block of the  $\hat{\vb{F}}$ has $n_m$ rows, where $n_m = \lceil N_m /p_c\rceil$. Using the standard matvec error result from~\cite{higham2002accuracy} gives
$$\|\delta\vb{v}_3\| \leq \|\hat{\vb{F}}\|\|\delta\vb{v}_2\| + \|\delta\hat{\vb{F}}\|\|\vb{v}_2\| + c_3\epsilon_3n_m\|\hat{\vb{F}}\|\|\vb{v}_2\|.$$
The $\hat{\vb{F}}^*$ case is identical except the $n_m$ factor is replaced with $n_d$, where $n_d = \lceil N_d / p_r\rceil$. The memory operations before and after the SBGEMV do not affect the error.

\paragraph{Batched IFFT} This phase is similar to the FFT; the error is $\delta\vb{v}_4 = \text{IFFT}(\delta\vb{v}_3) + (\text{error in IFFT}(\vb{v}_3))$. The norm of the IFFT operator in this case is $1/\sqrt{2N_t}$. Using the error result from~\cite{vanloan1992computational} gives
$$\|\delta \vb{v}_4\|\leq \frac{1}{\sqrt{2N_t}}\|\delta\vb{v}_3\| + \frac{c_4\epsilon_4}{\sqrt{2N_t}}\log_2(2N_t)\|\vb{v}_3\|.$$

\paragraph{Unpadding and Reduction} Unpadding is a memory operation that does not affect the error. The reduction for $\vb{F}$ matvecs is computed over each row in the 2D processor grid (i.e., over $p_c$ processors), and the reduction for $\vb{F}^*$ matvecs is computed over each column of the processor grid (i.e., over $p_r$ processes). For the $\vb{F}$ matvec, the error in this phase is $\delta \vb{v}_5 = (\text{sum of errors } \delta\vb{v}_4\text{ from each of the $p_c$ processes})  + (\text{error in reduction of } \vb{v}_4)$. Using the reduction error bound from~\cite{higham2002accuracy} gives
$$||\delta\vb{v}_5|| \leq \sum_{k=1}^{p_c} ||\delta\vb{v}_{4,k}|| + c_5\epsilon_5\log_2(p_c)\sum_{k=1}^{p_c} ||\vb{v}_{4,k}||,$$
where $\delta\vb{v}_{4,k}$ and $\vb{v}_{4,k}$ are the values of $\delta\vb{v}_{4}$ and $\vb{v}_{4}$ on process $k$, respectively. The $\vb{F}^*$ matvec result is identical except that $p_c$ is replaced by $p_r$.

Now, $\|\delta\vb{v}_5\|$ is the final \textit{absolute} error. However, the more important term to consider is the \textit{relative} error $\|\delta \vb{v}_5\|/\|\vb{v}_5\|$. Now, $\vb{v}_5 = \text{Reduce$($IFFT$($SBGEMV$($FFT$($Broadcast}(\vb{v}_0)))))$. Using this fact and recursively back-substituting into the error expressions for each phase, we arrive at the final result:
\begin{equation}
\begin{aligned}
    \frac{\|\delta\vb{v}_5\|}{\|\vb{v}_5\|} \leq \kappa(\hat{\vb{F}})\big[c_1\epsilon_1 &+ (c_F\epsilon_d + c_2\epsilon_2 + c_4\epsilon_4)\log_2(N_t)\\&+ c_3\epsilon_3n_m + c_5\epsilon_5\log_2(p_c)\big],\label{eq:error}
\end{aligned}
\end{equation}
for the $\vb{F}$ matvec. The $\vb{F}^*$ matvec result is identical except that $n_m$ is replaced by $n_d$ and $p_c$ is replaced by $p_r$. In~\Cref{eq:error}, $\kappa(\hat{\vb{F}})$ denotes the condition number of $\hat{\vb{F}}$~\cite{golub2013matrix,bottcher1999introduction}. From this analysis, it is seen that the dominant error term comes from the SBGEMV in Phase 3. This is to be expected, as it is the crux of the entire FFTMatvec algorithm. Furthermore, as the relative error scales with the condition number of $\hat{\vb{F}}$, care must be taken when computing with ill-conditioned matrices that can often appear in application contexts. The regularization used in inverse problem settings can help mitigate the conditioning~\cite{ghattas2021learning}. As the practical error propagation properties will depend heavily on the specific problem context, the Pareto front analysis in~\Cref{sec:mp} is a useful tool to determine the optimal mixed-precision configuration to use for a given problem.

The performance-portable mixed-precision FFTMatvec application is available open-source at \url{https://github.com/s769/FFTMatvec/}. In the next section, we present numerical results corresponding to the performance portability optimizations and the mixed-precision framework described here.

\section{Results}\label{sec:results}

In this section, we present the numerical results for performance portability and mixed-precision computation of the FFTMatvec application. Specifically, we will discuss the effect of the performance optimization on the (conjugate) transpose SBGEMV kernel presented in~\Cref{sec:perf-opt} and a Pareto front analysis of the dynamic mixed-precision framework presented in~\Cref{sec:mp}. In each case, we begin by describing the tests used to obtain the numerical results and then discuss the results themselves. 

\subsection{Performance-Portable FFTMatvec}
\subsubsection{Performance-Optimized (Conjugate) Transpose SBGEMV}
\begin{figure*}[htb]
    \centering
    \includegraphics[width=\textwidth]{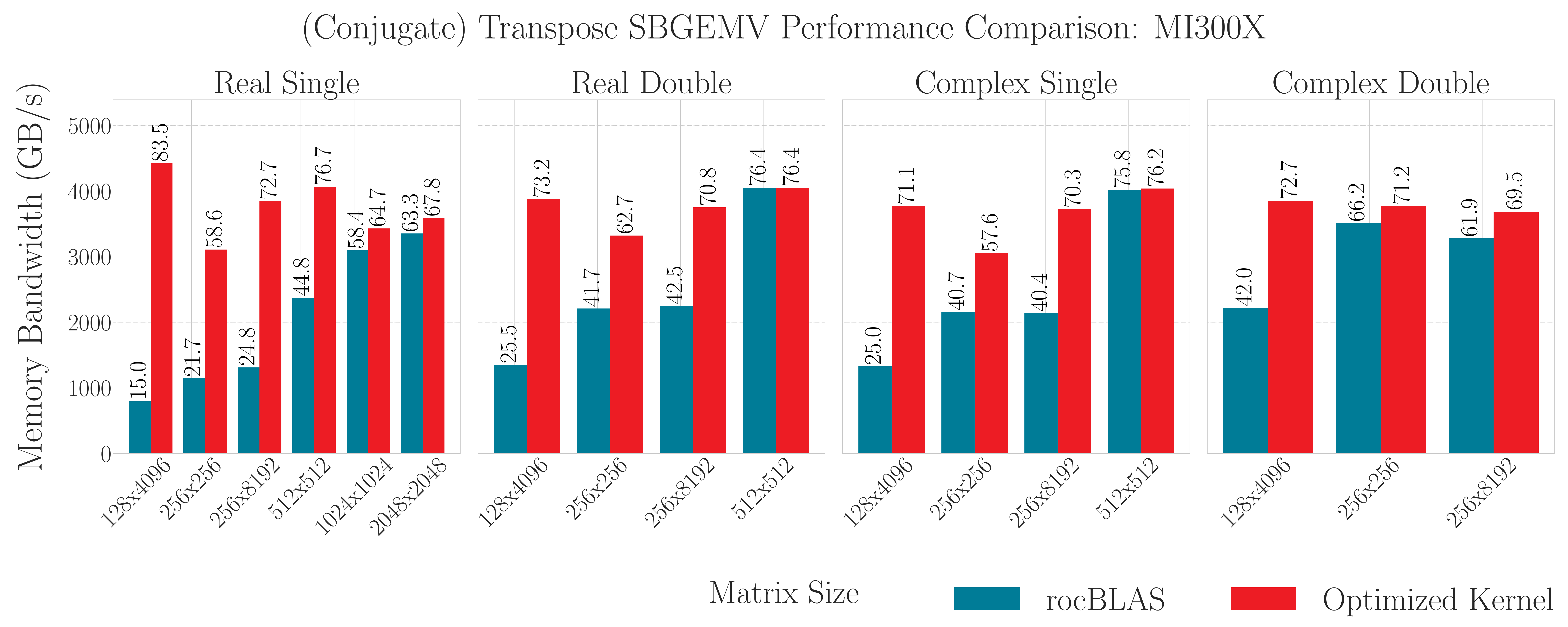}
    \caption{Performance comparison of rocBLAS vs. optimized implementation of strided batched GEMV (conjugate) transpose kernel for short, wide matrices ($m \leq n$) on an AMD Instinct\texttrademark \ MI300X GPU. Conjugate transpose is benchmarked for complex datatypes, and regular transpose is benchmarked for real datatypes. A batch size of 100 is used for all tests. Performance is measured by memory bandwidth as determined by the \texttt{rocblas-bench} benchmark. Bars are annotated with the percentage of peak memory bandwidth. The optimized kernel achieves greater relative performance on more skewed rectangular matrices than on square matrices and on lighter datatypes like real single than the heavy datatypes like double complex. See~\Cref{sec:perf-opt} for details on the optimized kernel implementation.}
    \label{fig:mi300x-rocblas-sbgemv-comp}
    \Description{Bar plots showing performance improvement of optimized SBGEMV kernel over the existing rocBLAS implementation for various datatypes and matrix sizes.}
\end{figure*}

\Cref{fig:mi300x-rocblas-sbgemv-comp} shows the results of the performance optimizations on the SBGEMV kernel described in~\Cref{sec:perf-opt}. The performance was measured on a single AMD Instinct\texttrademark \ MI300X GPU with ROCM version 6.4.1 using the \texttt{rocblas-bench} performance benchmarking tool. As the operation is memory-bound, performance is measured using the memory bandwidth metric.~\Cref{fig:mi300x-rocblas-sbgemv-comp} shows that the optimized kernel implementation achieves greater relative performance for smaller datatypes and matrices whose dimensions are more skewed (i.e., $m \ll n$) than for heavier datatypes and square matrices. For larger values of $m$, the existing rocBLAS implementation already performs well; the benchmarking results were also used to set the kernel transition points in the host launcher. Similar results were also observed for AMD Instinct\texttrademark \ MI250X GPUs---we have omitted the corresponding figure for brevity.

\subsubsection{Performance of FFTMatvec on AMD GPUs}
Once the conjugate transpose SBGEMV kernel was implemented, the full FFTMatvec application was benchmarked on the AMD Instinct\texttrademark \ MI250X,  MI300X, and MI355X GPUs. ROCm\texttrademark\ 6.4.1 was used for the  AMD Instinct MI250X and  MI300X tests, and ROCm 7.1.1 was used for the AMD Instinct MI355X tests. 

\begin{figure}
    \centering
    \includegraphics[width=\columnwidth]{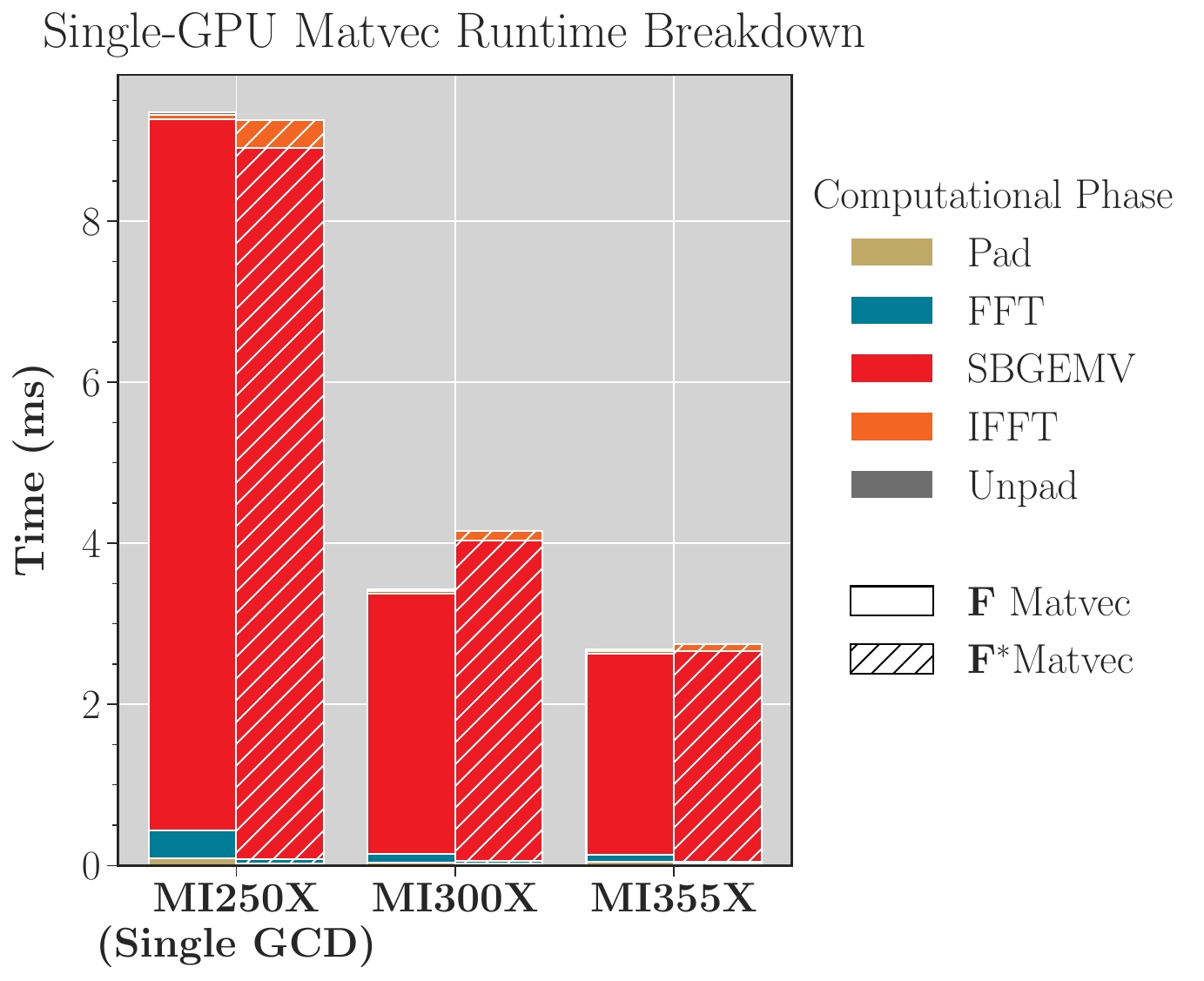}
    \caption{
    Runtime breakdown of FFTMatvec running on AMD Instinct\texttrademark \ MI250X (Single GCD),  MI300X, and  MI355X GPUs.
    The SBGEMV comprises the majority ($\sim$92\%) of the runtime. The left bar in each cluster shows the results for the $\blocktoep$ matvec, and the right bar shows the results for the $\blocktoep^*$ matvec. For all tests, $N_m = 5{,}000, N_d=100,$ and $N_t=1{,}000$. The observed trend in performance corresponds roughly to the peak memory bandwidths of the different GPUs.}
    \label{fig:single-gpu-runtimes}
    \Description{Clustered stacked bar plots showing the performance of FFTMatvec on various AMD GPUs.}
\end{figure}

The AMD Instinct MI250X module is composed of two Graphics Compute Dies (GCDs), each an independent GPU.  In single-GPU studies, only one GCD in an AMD Instinct \texttrademark \ MI250X module was used. For all tests, we used $N_m = 5{,}000, N_d=100,$ and $N_t=1{,}000$.
\Cref{fig:single-gpu-runtimes} shows the runtime breakdown for the $\blocktoep$ and $\blocktoep^*$ matvecs. The runtime is dominated by SBGEMV as expected, since this is the only operation that involves the entire matrix. 

The problem sizes tested here are characteristic of the ones found in the inverse problem setting, with $N_d \ll N_m$. Because SBGEMV is the primary single-GPU bottleneck, optimizing the transpose SBGEMV kernel ensures these results accurately represent FFTMatvec performance for various problem sizes. Detailed single-GPU performance studies over many different problem sizes can be found in~\cite{venkat2024fft}.

The observed trend in performance approximately correlates with the peak memory bandwidth of the different GPU architectures---1.6 TB/s $\to$ 5.3 TB/s $\to$ 8 TB/s going from AMD Instinct MI250X $\to$  MI300X $\to$  MI355X. This is expected as the entire application is memory-bound.

The $\blocktoep^*$ matvec on the AMD Instinct MI300X is slightly slower than the $\blocktoep$ matvec even when using the optimized conjugate transpose SBGEMV kernel. From profiling the application, we believe this is caused by the (non-transpose) GEMV kernel being extremely well-tuned on the AMD Instinct MI300X architecture for this problem size. For the other AMD GPU architectures, the $\blocktoep$ and $\blocktoep^*$ matvecs exhibit similar performance when the optimized conjugate transpose SBGEMV kernel is used. The SBGEMV kernels (both non-transpose and conjugate transpose) achieve approximately 70\% of the peak memory bandwidth on AMD Instinct MI250X and AMD Instinct MI300X. However, they only achieve approximately 35\% of the peak memory bandwidth on AMD Instinct MI355X. This is most likely due to the rocBLAS kernel parameters (grid and block sizes) and memory access patterns being optimized for AMD CDNA\texttrademark \ 2 and 3, with optimizations for CDNA 4 yet to be released. Notably, AMD CDNA 4 introduces increased LDS (shared memory) capacity and read-with-transpose instructions that could be leveraged to deliver much higher performance on the AMD Instinct MI355X GPUs.

\subsection{Pareto Front Analysis}
\subsubsection{Single GPU Results}
After the FFTMatvec application was benchmarked for performance on AMD GPUs, we ran the Pareto front analysis described in~\Cref{sec:mp}. 
We chose a relative error tolerance threshold of $10^{-7}$ and tested the 32 possible mixed-precision configurations\footnote{The 32 configurations result from the five computational phases, each of which can use single or double precision.} on AMD Instinct\texttrademark \ MI250X (single GCD),  MI300X, and  MI355X GPUs. As before, ROCm\texttrademark\ 6.4.1 was used for the  AMD Instinct MI250X and  MI300X tests, and ROCm 7.1.1 was used for the AMD Instinct MI355X tests. 

\begin{figure}
    \centering
    \includegraphics[width=\columnwidth]{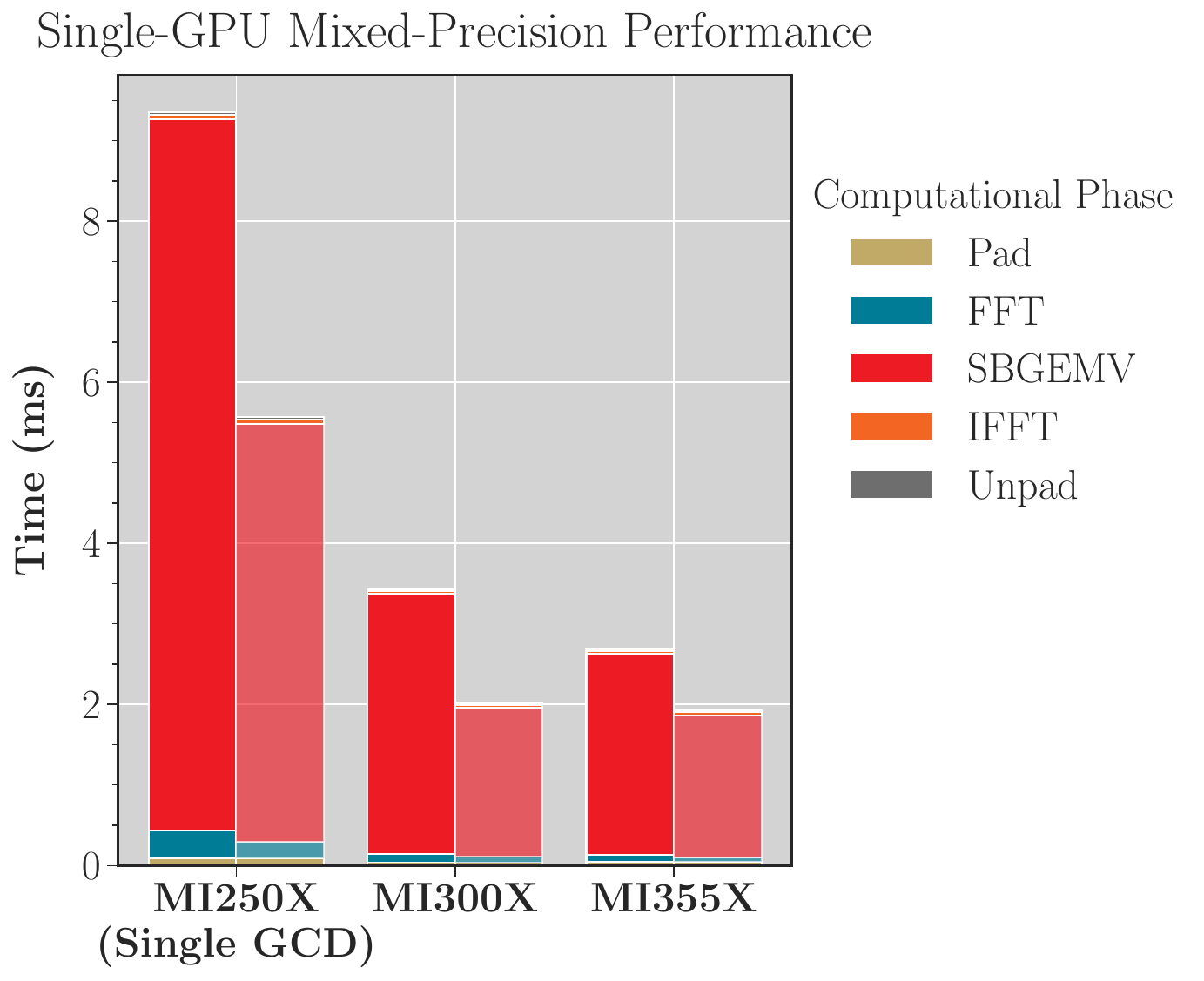}
    \caption{
    Double-precision vs. optimal mixed-precision configuration runtime breakdown of FFTMatvec ($\blocktoep$ matvec) running on AMD Instinct\texttrademark \ MI250X (Single GCD),  MI300X, and  MI355X GPUs. 
    The left bar in each cluster shows the baseline double-precision matvec, and the right bar shows the results for the matvec with optimal mixed-precision configuration for a relative error tolerance threshold of $10^{-7}$. Transparency is used to indicate a single-precision computational phase, while opacity indicates double precision. For all tests, $N_m = 5{,}000, N_d=100,$ and $N_t=1{,}000$.}
    \label{fig:mp-runtime-comparison}
    \Description{Clustered stacked bar plot showing optimal mixed-precision performance for FFTMatvec}
\end{figure}

For all tests, we again used $N_m = 5{,}000, N_d=100,$ and $N_t=1{,}000$. \Cref{fig:mp-runtime-comparison} shows the runtime breakdowns, speedups, and relative errors of the optimal mixed-precision configuration as compared to the baseline double-precision configuration for the $\blocktoep$ matvec ($\blocktoep^*$ results are similar). The optimal precision configuration for the $\blocktoep$ matvec on all three GPU architectures computes the FFT (of the input vector $\vb{m}$) and SBGEMV in single precision and all other phases in double precision. Similarly, the optimal precision configuration for the $\blocktoep^*$ matvec computes the SBGEMV and IFFT (of the vector $\vb{m}$, which is the output vector for $\blocktoep^*$ matvecs) in single precision and all other phases in double precision. This reflects the fact that the SBGEMV and FFT/IFFT of $\vb{m}$ together comprise nearly 97\% of the total runtime.

While computing the other phases in single precision can speed up those individual phases, the contribution to overall speedup is negligible. At the same time, such computations incur additional error. As a result, those configurations end up off the Pareto front.

Note that we initialized the matrices and vectors with double-precision floating point values that cannot be accurately represented as single-precision floating point numbers. This was done by setting mantissa bits in positions greater than 23 to one. Without this additional step, computing the broadcast in single precision would not incur any error, biasing the Pareto front analysis.

We observe 70\%-95\% speedups on the AMD Instinct MI250X and MI300X GPUs and a 40\% speedup on the AMD Instinct MI355X GPU. As observed previously, this can most likely be mitigated by optimizing rocBLAS kernels for AMD CDNA\texttrademark\ 4.

\subsubsection{Multi-GPU Results}
After benchmarking the mixed-precision framework for FFTMatvec on single GPUs, we performed scaling tests on the Oak Ridge Leadership Computing \textit{Frontier} supercomputer (\#2 on the Top500 as of June 2025\footnote{\url{https://top500.org/lists/top500/2025/06/}}) that is equipped with 9,472 nodes that each have eight AMD Instinct\texttrademark \ MI250X GPUs (counting a single GCD as a single GPU). Communication-aware partitioning was used to set the processor grid shape for FFTMatvec. One processor row was used when computing on 512 or fewer GPUs, eight processor rows were used for 1,024 and 2,048 GPUs, and 16 processor rows were used for 4,096 GPUs. The global problem size for $p$ GPUs was set to $N_m=5,000p$, $N_d=100$, and $N_t=1,000$.

\begin{figure}
    \centering
    \includegraphics[width=\columnwidth]{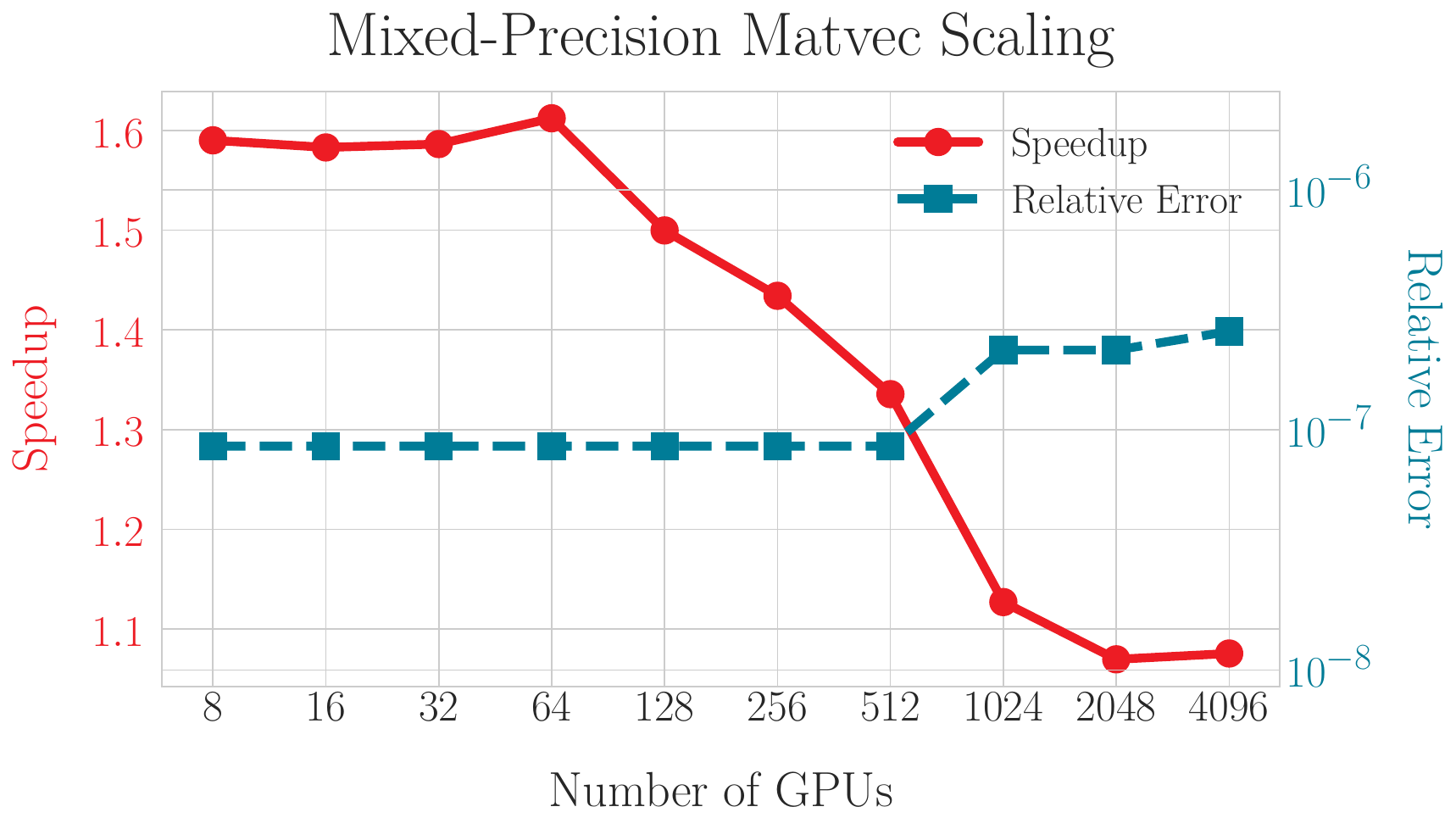}
    \caption{Speedups and relative errors of optimal mixed-precision configurations compared to the double-precision baseline when scaling from 8 to 4,096 GPUs on the \textit{Frontier} supercomputer ($\blocktoep$ matvec only; $\blocktoep^*$ results are similar). Communication-aware partitioning was used to select the optimal processor grid shape for each number of GPUs. The global problem size for $p$ GPUs was set to $N_m=5,000p$, $N_d=100$, and $N_t=1,000$. On 4,096 GPUs, a matvec with over 20 billion parameters ($N_mN_t$) is computed in $\sim 0.11s$. }
    \label{fig:mp-scaling}
    \Description{Line plots showing mixed-precision speedups and errors as FFTMatvec is scaled to 4,096 GPUs.}
\end{figure}

In the \textit{hipified} FFTMatvec code, the RCCL library is used for GPU-GPU communications. In order to achieve optimal communication performance on Frontier, the \texttt{open-ofi-plugin}\footnote{\url{https://github.com/HewlettPackard/open-ofi-xccl/commit/5338678a2da06f2374a25baa7ac4dac7ee3628c8}} along with the development branch of RCCL 2.22.3\footnote{\url{https://github.com/ROCm/rccl} --- commit \texttt{
e2c9f2f}} and ROCm\texttrademark\ 6.4.1 were used. Additionally, the GPU binding for MPI was set to the ``closest'' option, and the system environment was configured as in the OLCF documentation on best practices for RCCL.\footnote{\url{https://docs.olcf.ornl.gov/software/analytics/pytorch\_frontier.html\#environment-variables}; this page also contains instructions on how to build the \texttt{open-ofi-xccl} plugin.} The 32 mixed-precision configurations were run to determine the optimal configuration for each number of GPUs. \Cref{fig:mp-scaling} shows the speedups and relative errors of the optimal mixed-precision configuration for the $\blocktoep$ matvec as the number of GPUs increases. As the problem is scaled to more and more GPUs, communication costs dominate the runtime. Since the communication buffer sizes range from 0.8~MB (local data vector) to 40~MB (local parameter vector) while the network bandwidth is 100~GB/s, the communication is latency bound. As a result, communication in lower precision does not provide much speedup, but does increase the relative error in the result (see~\Cref{sec:error}). Thus, the Pareto front analysis shows that computing only the SBGEMV and FFT of the parameter vector $\vb{m}$ in single precision is the optimal strategy. 

Even as FFTMatvec is scaled to 4,096 GPUs, the relative error in the result remains under $10^{-6}$. The slight increase in relative error when running on more than 512 GPUs can be explained by looking at the relative error equation~\Cref{eq:error}. The dominant term in the error comes from the SBGEMV; this term is proportional to the \textit{local} parameter vector size $n_m = \lceil N_m / p_c\rceil$. As noted earlier, for the cases running on more than 512 GPUs, the optimal number of rows in the processor grid grows from 1 to 8 and then 16. As a result, $p_c$ becomes correspondingly smaller, increasing $n_m$. However, in~\Cref{eq:error}, the communication error term is proportional to $\log_2(p_c)$; this term decreases when we decrease $p_c$. The exact interplay between the two error terms is difficult to quantify theoretically without knowing the exact implementation details and algorithm-dependent constants. Nevertheless, the qualitative behavior of the relative error is justifiable by the preceding analysis. The numerical results also suggest that the error grows slowly when scaling to many thousands of GPUs.

Finally, it is relevant to note that the extreme-scale Bayesian inverse problem in~\cite{henneking2025bell} involving over one billion parameters and 600 sensors was solved using 512 GPUs, each with 80~GB of memory. This would correspond to 640 AMD Instinct MI250X GPUs (that each have 64~GB of memory). At that scale, the optimal mixed-precision configuration provides a $\sim$30\% speedup over the baseline double-precision computation. Additionally, the increased memory sizes of newer-generation GPUs---192~GB on the AMD Instinct\texttrademark \ MI300X and 288~GB on the AMD Instinct\texttrademark \ MI355X---mean that larger problems can fit on fewer numbers of GPUs, further reducing communication costs and increasing the overall speedup obtained from the mixed-precision computation. While a 30\% speedup might seem insignificant for a single matvec requiring a fraction of a second at baseline, as mentioned in~\Cref{rem:why}, it can result in significant time reductions in solutions of “outer-loop” problems that can require computing millions of these matvecs.

It is important to note that the key performance metric for FFTMatvec is time-to-solution rather than scalability. The global communication phases are an essential part of the FFTMatvec algorithm. This is in contrast to other applications such as PDE solvers, where the communication patterns are usually local~\cite{chan1994domain}. As a result, as FFTMatvec is scaled to many GPUs, its performance will be similar to that of global communication routines such as AllReduce. Communication-aware partitioning provided speedups of over $3\times$ when computing at 4,096 GPUs. Using this optimal partitioning scheme on 4,096 GPUs, a matvec with over 20 billion parameters ($N_mN_t$) is computed in $\sim 0.11$ seconds.

It is difficult to fully overlap communication with computation in FFTMatvec, as the computational Phases 2-4 rely on the results of the communication in Phase 1. When computing individual matvecs, the result of the reduction in Phase 5 is the final output; there is no further overlapping that can be done. However, when computing many matvecs in sequence and saving the results to file, the matvec calls can be overlapped with the host routines that generate input vectors and save output vectors. This process is used when computing dense operators that are relevant to solving Bayesian inverse problems in real time~\cite{henneking2025bell}.

\section{Conclusion}\label{sect:conclusion}

As GPU hardware, driven by the growing market for AI, continues to focus on improving lower precision (FP32 and below) performance, ``traditional'' scientific workloads are hard-pressed to adapt by leveraging mixed-precision algorithms. In addition, as the large supercomputing clusters used for scientific computing remain diversified in their choice of hardware vendors, performance portability becomes a critical part of many HPC workflows. In this paper, we presented a framework for performance portability via \texttt{hipify} on-the-fly
for an HPC application---FFTMatvec---that computes matrix-vector products with block-triangular Toeplitz matrices using an FFT-based, GPU-accelerated algorithm. The \texttt{hipify} on-the-fly
approach enabled the pure CUDA source code of FFTMatvec to be converted to HIP at compile time in order to run seamlessly on AMD GPUs. The portability framework, powered by \textit{hipify}, avoided code refactoring and issues with multiple source versions while keeping the user-facing code simple and readable. Performance optimizations for AMD GPUs were integrated directly into the open-source rocBLAS library, keeping the application code unchanged. The optimizations to the (conjugate) transpose SBGEMV kernel resulted in significant performance improvements over the existing rocBLAS implementation. 

In addition to performance portability, this work introduced a dynamic framework for using mixed precision in FFTMatvec. Through a Pareto front analysis, the optimal mixed-precision configuration for a given error tolerance was determined. Moreover, a theoretical analysis of the numerical error in the mixed-precision FFTMatvec was presented.
The entire application was benchmarked on AMD Instinct\texttrademark \ MI250X, AMD Instinct\texttrademark \ MI300X, and the newly launched AMD Instinct\texttrademark \ MI355X GPUs, showing excellent performance. Moreover, the mixed-precision framework was scaled to 4,096 GPUs on the \textit{Frontier} supercomputer and gave an approximate 30\% speedup over the baseline double-precision algorithm at 640 AMD Instinct MI250X GPUs---the amount it would take to solve a Bayesian inverse problem with over one billion parameters~\cite{henneking2025bell}. Thus, the mixed-precision framework provides considerable computational advantages for solving ``outer-loop'' problems such as that of optimal sensor placement at large scales.

The FFTMatvec application itself has been used to solve large scale Bayesian inference problems, specifically for tsunami early warning~\cite{henneking2025bell,henneking2025goal}. The algorithmic framework, however, is applicable to many other problems, including inverse problems for acoustic, electromagnetic, and elastic inverse scattering; source inversion for transport of atmospheric or subsurface hazardous agents; satellite inference of emissions; and treaty verification. In addition, block-triangular Toeplitz matrix-vector products appear in the contexts of multi-channel signal processing and vector-autoregressive-moving-average models in econometrics~\cite{kailath1980linear,simpkins2012system,lutkepohl2013introduction,saad2003iterative}. As a result, the FFTMatvec application has broad applicability; the mixed-precision and performance portability frameworks introduced here will enable FFTMatvec to better tackle these various problems.

\begin{acks}
This research used resources of the Oak Ridge Leadership Computing Facility at the Oak Ridge National Laboratory, which is supported by the Office of Science of the U.S. Department of Energy under Contract No. DE-AC05-00OR22725. This research was supported by DOD MURI grants FA9550-24-1-0327, DOE ASCR grant DE-SC0023171, and NSF grant DGE-2137420.
\end{acks}


\bibliographystyle{ACM-Reference-Format}

\appendix

\appendixAD

\section{Overview of Contributions and Artifacts}

\subsection{Paper's Main Contributions}

\begin{description}
    \item[$C_1$] Performance-portable implementation of FFT-based, GPU-accelerated matrix-vector products for block-triangular Toeplitz matrices (known as FFTMatvec). 
    \item[$C_2$] Optimized implementation of (conjugate) transpose GEMV kernel in rocBLAS.
    \item[$C_3$] Dynamic mixed-precision framework for FFTMatvec.
\end{description}

\subsection{Computational Artifacts}

\begin{description}
    \item[$A_1$] https://doi.org/10.5281/zenodo.17162841 or\\ https://github.com/s769/FFTMatvec/
    \item[$A_2$] https://github.com/ROCm/rocm-libraries/tree/develop/projects (no DOI available --- production software)
\end{description}

\begin{center}
\begin{tabular}{rll}
\toprule
Artifact ID  &  Contributions &  Related \\
             &  Supported     &  Paper Elements \\
\midrule
$A_1$   &  $C_1$, $C_3$ & Figures 2-4 \\
\midrule
$A_2$   &  $C_2$ & Figure 1 \\
\bottomrule
\end{tabular}
\end{center}

\section{Artifact Identification}

\newartifact

\artrel

This artifact contains the FFTMatvec code that is the basis for the algorithms developed in the paper. The FFTMatvec application computes matrix-vector products with block-triangular Toeplitz matrices. The artifact $A_1$ specifically contains the performance portable, mixed-precision version of FFTMatvec.
\artexp

The FFTMatvec application in artifact $A_1$ can run on AMD GPUs. Running with the configurations reported below and in the paper should reproduce the results in Figures 2-4.

\arttime

The expected time to build the code for FFTMatvec is 30 minutes. The expected time to run single GPU results is 15-20 minutes. The expected time to run multi-GPU scaling results can be several days, depending on the job queue time. The expected time to analyze the results and generate plots is 1-2 hours.

\artin

\artinpart{Hardware}

For single GPU results, AMD Instinct\texttrademark \ MI250X, MI300X, and MI355X GPUs are required. For multi-GPU results, the AMD Instinct\texttrademark \ MI250X GPUs are required.

\artinpart{Software}

The FFTMatvec application requires the ROCm\texttrademark \ software development kit.  ROCm\texttrademark \ v6.4.1 should work for the AMD Instinct\texttrademark \ MI250X and MI300X GPUs, and ROCm\texttrademark \ v7.1.1 should work for the AMD Instinct\texttrademark \ MI355X GPUs. These can be found at \url{https://github.com/ROCm/ROCm}.

In addition, FFTMatvec also requires MPI (\url{https://www.open-mpi.org/}) and HDF5-parallel (\url{https://github.com/HDFGroup/hdf5/blob/develop/release_docs/INSTALL_parallel}).

\artinpart{Datasets / Inputs}

No datasets are required for this artifact.

\artinpart{Installation and Deployment}

An AMD GPU compiler is required to compile the FFTMatvec software for AMD GPUs. This is most often the \texttt{amd-clang++} compiler. The version shipped with ROCm\texttrademark \ v6.4.1 should work for the AMD Instinct\texttrademark \ MI250X and MI300X GPUs, and the version shipped with ROCm\texttrademark \ v7.1.1 should work for the AMD Instinct\texttrademark \ MI355X GPUs. 

\artcomp

The workflow begins with running the FFTMatvec application without a specified precision configuration to determine the baseline double-precision performance. Then, mixed-precision configurations are run, and the error is calculated with respect to the double-precision output. The mixed-precision configuration that provides the greatest speedup for a desired error tolerance is chosen. The application reports the average timing results over 100 repetitions.

\artout

The FFTMatvec application produces timing results for all the different computational phases. These results are used to generate the plots in Figures 2-4.
\newartifact

\artrel

This artifact is the ROCm\texttrademark \ rocBLAS library. It is included since the optimized transpose GEMV kernel that was described in the paper is now merged into the rocBLAS development branch. 
\artexp

The rocBLAS GitHub repository can be used to verify the results of the optimized transpose GEMV kernel in Figure 1.

\arttime

The expected time to build the code for rocBLAS is approximately 2 hours per version. The expected time to run the benchmarking tests is 10 minutes per version. The expected time for analysis to reproduce the results in Figure 1 is 1~hr.

\artin

\artinpart{Hardware}

At least one AMD Instinct\texttrademark \ MI300X is required to reproduce the results in Figure 1. 
\artinpart{Software}

The FFTMatvec application requires the ROCm\texttrademark \ software development kit.  ROCm\texttrademark \ v6.4.1 should work for the AMD Instinct\texttrademark \ MI300X GPU. rocBLAS needs to be built from source. 

\artinpart{Datasets / Inputs}

The \texttt{rocblas-bench} executable can take a \texttt{yaml} file with the problem sizes and datatypes as an input. The details of this file are given in the AE appendix below.

\artinpart{Installation and Deployment}

An AMD GPU compiler is required to compile the FFTMatvec software for AMD GPUs. This is most often the \texttt{amd-clang++} compiler. The version shipped with ROCm\texttrademark \ v6.4.1 should work for the AMD Instinct\texttrademark \ MI300X GPUs.

\artcomp

First, the two versions of the rocBLAS library (with clients enabled) are built. Then, the \texttt{yaml} file with the problem configurations for Figure 1 is created. Next, the \texttt{rocblas-bench} executable is run with the \texttt{yaml} file as input. Finally, the outputs of the \texttt{rocblas-bench} from both rocBLAS versions are compared to reproduce Figure 1. 

\artout

The \texttt{rocblas-bench} output (\texttt{rocblas-GB/s}) is shown in Figure 1. The application averages over a set number of repetitions.

\appendixAE

\arteval{1}
\artin

\artcomp

The FFTMatvec application can be built by cloning the repository at \url{https://github.com/s769/FFTMatvec/}. The \texttt{README.md} file in the repository has build instructions. 

Sometimes, the \texttt{CMAKE\_PREFIX\_PATH} must be set properly for the application and tests to build. If the application (\texttt{fft\_matvec}) builds but the tests do not build, all the results in the paper can still be generated. To do a manual test of the \texttt{fft\_matvec} executable, simply run \texttt{./fft\_matvec -t} from the build directory. If the tests do build, they can be run with \texttt{ctest}.

The FFTMatvec README file has detailed instructions on how to run the executable. The main arguments to set are \texttt{-nm 5000 -nd 100 -Nt 1000}. For running mixed-precision tests, also set \texttt{-rand}. The 32 possible test configurations are set with \texttt{-prec xxxxx} where each \texttt{x} can take the value \texttt{d} or \texttt{s}. The \texttt{-raw} option can be used to make the output more easily parsed by other scripts. If it is not set, the output is more human-readable. The \texttt{-s <directory>} option can be used to save the output vectors in the given directory. This is useful for comparing mixed-precision and double-precision outputs. 

For multi-GPU tests, use \texttt{mpirun -n <num-processes> ./fft\_matvec <args>}. It is enough to not pass anything for \texttt{-pr} and \texttt{-pc}; they will be set automatically to the values used in the paper. See Section 4.2.2 for how to optimally configure RCCL on \textit{Frontier}.

\artout

The output of \texttt{fft\_matvec} has the timing results of each portion of the computation. For the figures in the paper, some timing results may need to be combined to reflect the computational phases outlined in the paper. The SBGEMV time includes the SOTI-to-TOSI and TOSI-to-SOTI times. The first three lines of timing output show the setup, total time, and cleanup times; these are not used in the paper. The next three lines of timing output show the times for the $\mathbf{F}$ matvec. These are the mean, min, and max times among all processes, respectively. The last three lines of timing output show the same results for the $\mathbf{F}^*$ matvec.

The double-precision, single-GPU times are used to generate Figure 2. Running with a precision configuration of \texttt{-prec dssdd} should reproduce the results in Figure 3. Running with the number of GPUs found in Figure 4, using a precision configuration of \texttt{-prec dssdd} for fewer than 512 GPUs, and \texttt{-prec dssds} for 512 or more GPUs, should reproduce the results in Figure 4. Our experiments were run on the OLCF \textit{Frontier} machine.

\arteval{2}
\artin
The instructions for cloning the rocBLAS library from the \texttt{rocm-libraries} monorepo are found at 
\url{https://github.com/ROCm/rocm-libraries/blob/develop/CONTRIBUTING.md}. A sparse checkout of rocBLAS should be sufficient. By checking out a commit dated between June 1, 2025, and August 1, 2025 (e.g., \texttt{cf7df1d}), a version of rocBLAS without the optimized kernel can be obtained. By checking out commit \texttt{dd7ea70} or \texttt{12486d2} (slightly updated), a version with the optimized kernel is obtained. 

Then, use the \texttt{install.sh} script to build rocBLAS. The options \texttt{-c -n} should be used. The \texttt{-d} can be used to automatically install dependencies but requires administrator privileges. Otherwise, the dependencies \texttt{libdrm} and \texttt{gtest} need to be installed and set in \texttt{CMAKE\_PREFIX\_PATH}. Also, the \texttt{-a gfx942} is set for Figure 1.
\artcomp
The \texttt{rocblas-bench} executable is used to run the tests for Figure 1. The executable is found in \texttt{build/release/clients/staging/}. It accepts a \texttt{yaml} file that specifies the different test configurations to run. This \texttt{yaml} file contains entries such as 
\begin{lstlisting}
    - {M: 128, N: 4096, alpha: 1.0, batch_count: 100, beta: 0.0, cold_iters: 2, incx: 1, incy: 1, iters: 10, lda: 128, rocblas_function: rocblas_sgemv_strided_batched, stride_a: 524288, stride_x: 4096, stride_y: 128, transA: T}
\end{lstlisting}
To run the tests to generate Figure 1, the \texttt{M, N, lda, stride\_a, stride\_x, stride\_y, transA} and \texttt{rocblas\_function} parameters need to be set. \texttt{M = lda = stride\_y} is the number of rows in each matrix, \texttt{N = stride\_x} is the number of columns in each matrix, \texttt{stride\_a = M*N}. The \texttt{transA} parameter is set to \texttt{T} for real datatypes and \texttt{H} for complex datatypes. The \texttt{rocblas\_function} is set to \texttt{rocblas\_xgemv\_strided\_batched}, where \texttt{x} is \texttt{s} (real single), \texttt{d} (real double), \texttt{c} (complex single), or \texttt{z} (complex double). 

A single \texttt{yaml} file containing all the problem sizes and datatypes reported in Figure 1 can be made and saved as \texttt{conf.yaml}

The \texttt{rocblas-bench} executable is run with\\
\texttt{./rocblas-bench --yaml conf.yaml > out.txt}. 
\artout

After running \texttt{rocblas-bench} on both the optimized and unoptimized rocBLAS versions, the \texttt{out.txt} of each version contains a CSV file with the outputs of each test configuration. From these, the values corresponding to the \texttt{rocblas-GB/s} for each test case can be plotted for the two rocBLAS versions to reproduce Figure 1.

\end{document}